\documentclass[nofootinbib,a4paper]{revtex4}
\usepackage{graphicx}
\usepackage{fancyhdr}
\usepackage{amsmath}
\usepackage{mathrsfs}
\def\sla#1{\rlap{\kern .22em /}#1}%文字にスラッシュを重ねる$\sla{}$
\newcommand{\lromn}[1]{\uppercase\expandafter{\romannumeral#1}}

\newcommand{\MNS}{{\text{MNS}}}
\newcommand{\eV}{{\text{eV}}}

\newcommand{\GeV}{{\text{GeV}}}
\newcommand{\TeV}{{\text{TeV}}}
\newcommand{\BR}{\text{BR}}

\newcommand{\BL}{{\text{B$-$L}}}
\newcommand{\DM}{{\text{DM}}}

\newcommand{\U}{{\text{U}}}

\newcommand{\SI}{{\text{SI}}}

\usepackage{color}

\begin{document}

\title{Neutrino Mass and Dark Matter from Gauged B$-$L Breaking
\footnote{This talk is based on Ref.~\cite{Kanemura:2014rpa}. }
}


\author{Toshinori Matsui}
\email{matsui@jodo.sci.u-toyama.ac.jp}
\affiliation{Department of Physics, University of Toyama, Toyama 930-8555, Japan}

\begin{abstract}
We discuss a new radiative seesaw model with the gauged B$-$L symmetry which is spontaneously broken. 
We improve the previous model by using the anomaly-free condition without introducing too many fermions. 
In our model, dark matter, tiny neutrino masses and neutrino oscillation data can be explained simultaneously, 
assuming the B$-$L symmetry breaking at the TeV scale. 
%assuming the spontaneous breaking of B$-$L symmetry at the TeV scale. 
\end{abstract}

\maketitle

\thispagestyle{fancy}


%%%%%%%%%%%%%%%%%%%%%%%%%%%%%%%%%%
%%%%%%%%%%%%%%%%%%%%%%%%%%%%%%%%%%
%%%%%%%%%%%%%%%%%%%%%%%%%%%%%%%%%%
%%%%%%%%%%%%%%%%%%%%%%%%%%%%%%%%%%
\section{Introduction}

%%1
%1
The neutrino oscillation data~\cite{Aharmim:2011vm, Abe:2014ugx, An:2013zwz} have shown us neutrinos have tiny masses. 
%2
If $\nu_R^{}$ are introduced to the standard model of particle physics~(SM), there are two possible mass terms for neutrinos~(See e.g., Ref.~\cite{Ref:seesaw}), the Dirac type $\overline{\nu_L}\nu_R$ and the Majorana type $\overline{(\nu_R^{})^c} \nu_R^{}$.
%3
In radiative seesaw models~(See e.g., Refs.~\cite{Ref:KNT, Ref:Ma, Ref:AKS, Aoki:2011yk, Kanemura:2012rj, Ref:KNS, Kanemura:2013qva}), an {\it ad hoc} unbroken $Z_2$ symmetry forbids generating neutrino masses at the tree level and explains the dark matter~(DM) stability. 
%4
A model in Ref.~\cite{Ref:KNS} was constructed such that the breaking of the $\U(1)_\BL$ gauge symmetry gives a residual symmetry for the DM stability and the Majorana neutrino mass of $\nu_R^{}$. 
%5
However, the anomaly cancelation for the $\U(1)_\BL$ gauge symmetry requires to introduce more additional fermions except for particles for the radiative neutrino mass.

%%2
%1
In this talk, we propose a new model which is an improved version of the model in Ref.~\cite{Ref:KNS} from the view point of the anomaly cancellation. 
%2
With appropriate $\U(1)_\BL$  charge assignments, there exists an unbroken global $\U(1)$ symmetry even after the breakdown of the $\U(1)_\BL$ symmetry. 
%3
The global $\U(1)$ symmetry stabilizes the DM, so that we hereafter call it $\U(1)_\DM$. 
%4
In our work, the DM candidate is a new scalar boson. 
%5
Furthermore, the Dirac mass term of neutrinos is radiatively generated at the one-loop level due to the quantum effect of the new particles. 
%6
Tiny neutrino masses are explained by the two-loop diagrams with a Type-I-Seesaw-like mechanism. 
%7
We find that the model can satisfy current data from the neutrino oscillation, the lepton flavor violation~(LFV), the relic abundance and the direct search for the DM, and the LHC experiment.

%%%%%%%%%%%%%%%%%%%%%%%%%%%%%%%%%%
%%%%%%%%%%%%%%%%%%%%%%%%%%%%%%%%%%
%%%%%%%%%%%%%%%%%%%%%%%%%%%%%%%%%%
%%%%%%%%%%%%%%%%%%%%%%%%%%%%%%%%%%
\section{Model}

%%1
%1
We introduce new particles which listed in Table~\ref{tab:particle}. 
%2
We determine assignment of $\U(1)_\BL$ charges from conditions for cancellation of the $[\U(1)_\BL]\times[\text{gravity}]^2$ and $[\U(1)_\BL]^3$ anomalies; 
%%----- eq: Anomaly Cancellation >>>>>
\begin{eqnarray}
3
-\frac{1}{\,3\,} N_{\nu_R^{}}
-\frac{2}{\,3\,} N_\psi
= 0 , \quad
%%-----------------
3
-\frac{1}{27} N_{\nu_R^{}}
+\left( -2x^2 - \frac{4}{\,3\,} x - \frac{8}{27} \right) N_\psi
= 0, 
\end{eqnarray}
%%<<<<< eq: Anomaly Cancellation -----
where $N_\psi$ is the number of $\psi_{Ri}^{}$ (the same as the number of $\psi_{Li}^{}$), and $N_{\nu_R^{}}$ is the number of $\nu_{Ra}^{}$. 

%1
There are four solutions as presented in Table~\ref{tab:BL-charge}. 
%2
Except for Case~III, the $\U(1)_\BL$ charges of some new particles are irrational numbers while the $\U(1)_\BL$ symmetry is spontaneously broken by the vacuum expectation value~(VEV) of $\sigma^0$ whose $\U(1)_\BL$ charge is a rational number. 
%3
Therefore, the irrational charges are conserved, and the lightest particle with an irrational $\U(1)_\BL$ charge becomes stable so that the particle can be regarded as a DM candidate. 
%4
In this talk, we take Case~IV as an example.
%
%---------------------
\begin{table}[t]
  \begin{center}
    \begin{tabular}{c}
      % 1
      \begin{minipage}[b]{0.5\hsize}
      {\renewcommand\arraystretch{1.17}
        \begin{center}\makeatletter\def\@captype{table}\makeatother
\caption{Particle contents in this model.
 Indices $i$ and $a$ run from $1$ to $N_\psi$ and from $1$ to $N_{\nu_R}$, respectively.}
\begin{tabular}
{|c|@{\vrule width 1.8pt\ } c|c|c|c|c|c|}
   \hline
     &$\sigma^0$&$(\nu_R)_a$&$(\psi_L)_i$&$(\psi_R)_i$&$\eta$&$s^0$
    \\ \noalign{\hrule height 1.8pt}
    SU(2)$_{\rm I}$&{\bf 1}&{\bf 1}&{\bf 1}&{\bf 1}&{\bf 2}&{\bf 1}
    \\ \hline
    U(1)$_{\rm Y}$&$0$&$0$&$0$&$0$&$1/2$&$0$
    \\ \hline
    U(1)$_{B-L}$&$2/3$&$-1/3$&$x+2/3$&$x$&$x+1$&$x+1$
    \\ \hline
        Spin&0&1/2&1/2&1/2&0&0
    \\ \hline
\end{tabular}
\label{tab:particle}
        \end{center}
        }
      \end{minipage}
\qquad
      % 2
      \begin{minipage}[b]{0.4\hsize}
      {\renewcommand\arraystretch{1.6}
        \begin{center}\makeatletter\def\@captype{table}\makeatother
\caption{Sets of $N_\psi$, $N_{\nu_R^{}}$ and $x$, for which the $\U(1)_\BL$ gauge symmetry is free from anomaly. }
\begin{tabular}{|c|@{\vrule width 1.8pt\ }c|c|c|c|}
\hline
 {}
 & Case I
 & Case II
 & Case III
 & Case IV
    \\ \noalign{\hrule height 1.8pt}
 $N_\psi$
 & $1$
 & $2$
 & $3$
 & $4$
\\\hline
 $N_{\nu_R^{}}$
 & $7$
 & $5$
 & $3$
 & $1$
\\\hline
 $x$
 & $\frac{ 2\sqrt{3}-1 }{3}$
 & $\frac{ \sqrt{6}-1 }{3}$
 & $\frac{1}{\,3\,}$
 & $\frac{ \sqrt{3}-1 }{3}$
 \\\hline
\end{tabular}
\label{tab:BL-charge}
          \end{center}
          }
\end{minipage}
\end{tabular}
\end{center}
\end{table}
%---------------------

%%2
%1
 In addition to the SM one, the new Yukawa interactions are given by
 %%----- eq: Yukawa >>>>>
\begin{eqnarray}
{\cal L}_{\text{Y}}
=
-(y_R^{})_i\,\overline{ (\nu_R^{})_i }\, (\nu_R^{})^c_i\, (\sigma^0)^\ast
-(y_{\psi}^{})_i\, \overline{ (\psi_R)_i }\, (\psi_L)_i\, (\sigma^0)^\ast
-h_{ij}\,\overline{ (\psi_L)_i }\,  (\nu_R^{})_j \, s^0
-f_{\ell i}\,\overline{ (L_L)_\ell }\, (\psi_R)_i\, \tilde{\eta}+{\rm h.c.},\, 
 \label{eq:Yukawa}
\end{eqnarray}
%%<<<<< eq: Yukawa -----
where  $\tilde{\eta}\equiv i\sigma_2\, \eta^\ast$. 
 %2
 The scalar potential in our model is the same as that in the previous model~\cite{Ref:KNS}:
\begin{eqnarray}
V%(\Phi,\sigma,\eta,s)
 =
 -\mu_{\phi}^2\Phi^{\dagger} \Phi
 +
 \mu_s^2 |s^0|^2
 +
 \mu_{\eta}^2\eta^{\dagger} \eta
 -
 \mu_{\sigma}^2 |\sigma^0|^2
 +
\mu_3^{}\, (s^0\, \eta^{\dagger}\, \Phi+{\rm h.c.})
 +
 \lambda_\phi \left(\Phi^{\dagger} \Phi\right)^2
 +
 \lambda_s |s^0|^4
 +
 \lambda_\eta \left(\eta^{\dagger} \eta\right)^2
 +
 \lambda_\sigma |\sigma^0|^4  &&
  \nonumber\\
+
 \lambda_{s\sigma} |s^0|^2 |\sigma^0|^2
 +
 \lambda_{s\eta} |s^0|^2 \eta^{\dagger} \eta
 +
 \lambda_{s\phi} |s^0|^2 \Phi^{\dagger} \Phi
 +
 \lambda_{\sigma\eta} |\sigma^0|^2 \eta^{\dagger} \eta
 +
 \lambda_{\sigma\phi} |\sigma^0|^2 \Phi^{\dagger} \Phi
  +
 \lambda_{\phi\phi} (\eta^{\dagger}\eta) (\Phi^{\dagger} \Phi)
 +
 \lambda_{\eta\phi} (\eta^{\dagger} \Phi) (\Phi^{\dagger}\eta). \,&&
 \label{eq:V}
 \end{eqnarray}

%%3
%1
Neutral scalar fields are given by $\phi^0=\frac{1}{\sqrt{2}}(\phi^0_r+iz_\phi), \sigma^0=\frac{1}{\sqrt{2}}(\sigma^0_r+iz_\sigma), \eta^0=\frac{1}{\sqrt{2}}(\eta^0_r+i\eta^0_i), s^0=\frac{1}{\sqrt{2}}(s^0_r+is^0_i)$. 
%2
Two scalar fields $\phi^0$ and $\sigma^0$ obtain VEVs $v_\phi^{}$~[$= \sqrt{2}\, \langle \phi^0 \rangle = 246\,\GeV$] and $v_\sigma^{}$~[$= \sqrt{2}\, \langle \sigma^0 \rangle$].
%3
The VEV $v_\sigma^{}$ provides a mass of the $\U(1)_\BL$ gauge boson $Z^\prime$
as $m_{Z^\prime}^{} = (2/3) g_\BL^{} v_\sigma^{}$, where $g_\BL^{}$ is the $\U(1)_\BL$ gauge coupling constant.
%4
After the gauge symmetry breaking with $v_\phi^{}$ and $v_\sigma^{}$, we can confirm in Eqs.~\eqref{eq:Yukawa} and \eqref{eq:V} that there is a residual global $\U(1)_\DM$ symmetry, for which irrational $\U(1)_\BL$-charged particles ($\eta$, $s^0$, $\psi_{Li}^{}$, and $\psi_{Ri}^{}$) have the same $\U(1)_\DM$-charge while the other particles are neutral.

%%4
%1
Two CP-even scalar particles $h^0$ and $H^0$ are obtained by $\phi^0$-$\sigma^0$ mixing as $\sin{2\theta_0}=\frac{ 2\lambda_{\sigma\phi} v_\phi^{} v_\sigma }{ m_{H^0}^2 - m_{h^0}^2 }$. 
%2
Two neutral complex scalars $\eta^0$ and $s^0$ are obtained by $\eta^0$-$s^0$ mixing as $\sin{2\theta_0^\prime}=\frac{ \sqrt{2} \mu_3^{} v_\phi^{} }{ m_{{\mathcal H}_2^0}^2 - m_{{\mathcal H}_1^0}^2 }$. 
%3
Scalar masses are given by 
%%%%%%%%%%%%%%
\begin{eqnarray}
 m_{{h^0}, {H^0}}^2
=
 \lambda_\phi v_\phi^2 + \lambda_\sigma v_\sigma^2
 \mp \sqrt{
   \left(
    \lambda_\phi v_\phi^2 - \lambda_\sigma v_\sigma^2
   \right)^2
   + \lambda_{\sigma\phi}^2 v_\phi^2 v_\sigma^2 }, \,
%
%%%%%%%%%%%%%%
 m_{{{\mathcal H}_1^0}, {{\mathcal H}_2^0}}^2
=
 \frac{1}{2}
 \left(
  m_\eta^2 + m_s^2
  \mp\sqrt{\left( m_\eta^2 - m_s^2 \right)^2 + 2 \mu_3^2 v_\phi^2}
 \right), \,
\end{eqnarray}
%%%%%%%%%%%%%%
where $m_{\eta}^2
= \mu_{\eta}^2
 + \left( \lambda_{\phi\phi} + \lambda_{\eta\phi} \right) v_{\phi}^2/2
 + \lambda_{\sigma\eta} v_{\sigma}^2/2$, 
 $m_s^2
= \mu_s^2 + \lambda_{s\phi} v_{\phi}^2/2 + \lambda_{s\sigma} v_{\sigma}^2/2$. 
%
%4
The mass of the charged scalar~$\eta^{\pm}$ is $m_{\eta^{\pm}}^2 = m_\eta^2 - \lambda_{\eta\phi} \, v_{\sigma}^2 / 2$. 
Nambu-Goldstone bosons $z_\phi^{}$ and $z_\sigma^{}$ are absorbed by $Z$ and $Z^\prime$ bosons, respectively.

%%%%%%%%%%%%%%%%%%%%%%%%%%%%%%%%%%
%%%%%%%%%%%%%%%%%%%%%%%%%%%%%%%%%%
%%%%%%%%%%%%%%%%%%%%%%%%%%%%%%%%%%
%%%%%%%%%%%%%%%%%%%%%%%%%%%%%%%%%%
\section{Phenomenology}

%%%%%%%%%%%%%%%%%%%%%%%%%%%%%%%%%%
\subsection{Neutrino masses}

\begin{figure}[t]
  \begin{center}
    \begin{tabular}{c}
      % 1
      \begin{minipage}{0.5\hsize}
        \begin{center}
          \includegraphics[clip, width=6cm]{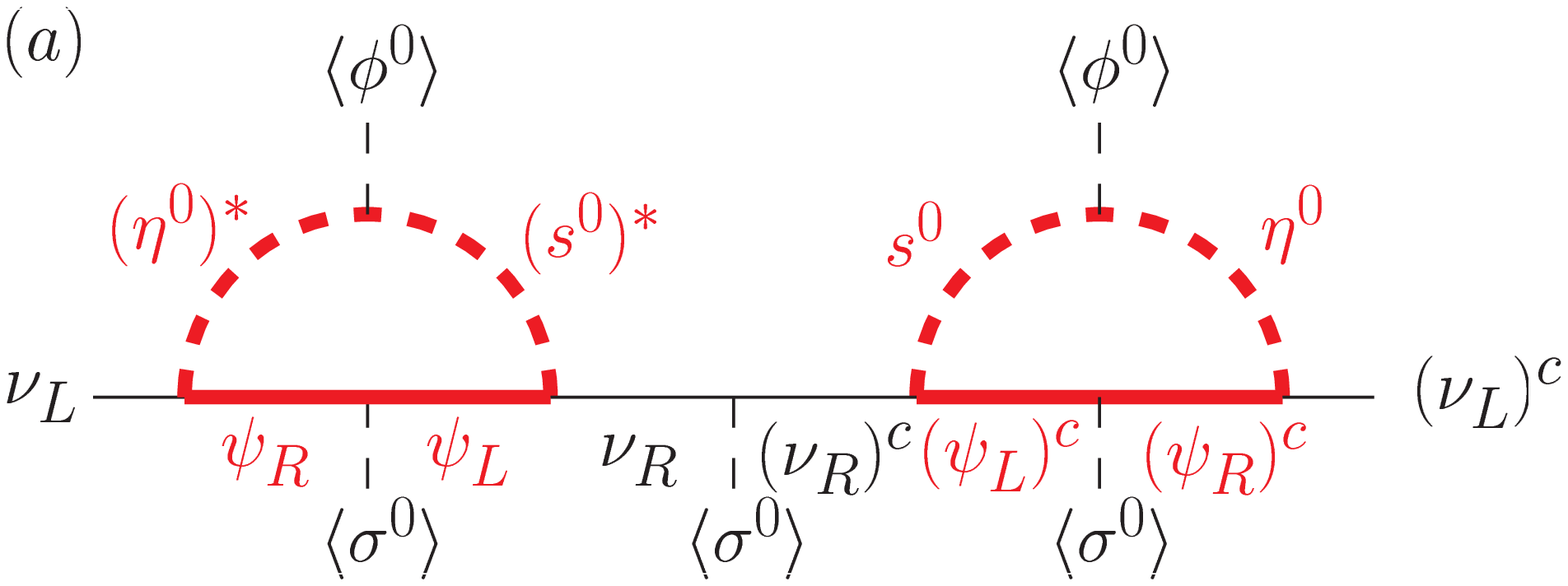}
        \end{center}
      \end{minipage}
      % 2
      \begin{minipage}{0.5\hsize}
        \begin{center}
          \includegraphics[clip, width=4.1cm]{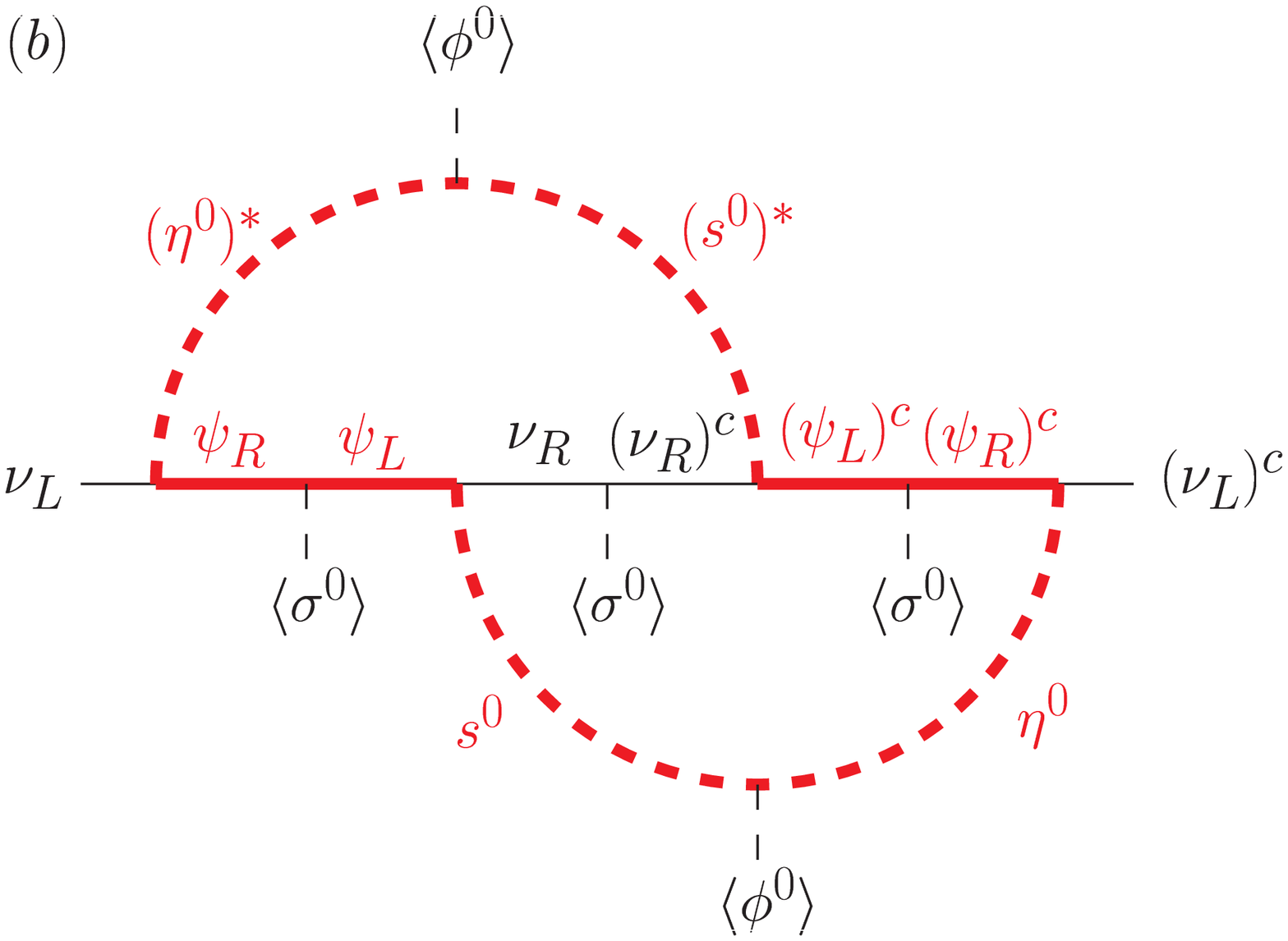}
        \end{center}
      \end{minipage}
    \end{tabular}
    \caption{ Two-loop diagrams for tiny neutrino masses in this model. }
    \label{Fig:two-loop}
  \end{center}
\end{figure}

%1
Tiny neutrino masses are generated by two-loop diagrams in Fig.~\ref{Fig:two-loop}~\cite{Ref:KNS}.
%2
The mass matrix $m_\nu^{}$ is expressed in the flavor basis as
%--------------------
\begin{eqnarray}
\left( m_{\nu} \right)_{\ell\ell'}
&=&
% \left(\frac{1}{16\pi^2}\right)^2
  \sum_{i,j,a}
  f_{\ell i}\,
  h_{ia}\, (m_R^{})_a\, (h^T)_{aj}\,
  (f^T)_{j\ell'}
  \Bigl[
   \left( I_1 \right)_{ija}
   + \left( I_2 \right)_{ija}
  \Bigr]
  /(16\pi^2)^2
  ,
 \label{eq:neutrinomass}
\end{eqnarray}
%-----------------------------
where explicit formulas of $\left( I_1 \right)_{ija}$ and $\left( I_2 \right)_{ija}$ are shown in Ref.~\cite{Kanemura:2014rpa}. 
%3
 The neutrino mass matrix $(m_\nu)_{\ell\ell^\prime}^{}$ is diagonalized by a unitary matrix $U_\MNS$, the so-called Maki-Nakagawa-Sakata~(MNS) matrix~\cite{Maki:1962mu}, as $U_\MNS^\dagger\, m_\nu\, U_\MNS^\ast
=
 \text{diag}( m_1 e^{i\alpha_1} ,\,
 m_2 e^{i\alpha_2} ,\, m_3 e^{i\alpha_3} )$. 
 %4
 We take $m_i$~($i= 1\text{-}3$) to be real and positive values. 
 %5
 Two differences of three phases $\alpha_i$ are physical Majorana phases. %~\cite{Ref:M-Phase}
% The MNS matrix can be parametrized as
%%----- eq: MNS >>>>>\begin{eqnarray}U_\MNS=
% \begin{pmatrix} 1 & 0 & 0\\ 0 & c_{23} & s_{23}\\ 0 & -s_{23} & c_{23} \end{pmatrix}
%\begin{pmatrix} c_{13} & 0 & s_{13} e^{-i\delta}\\ 0 & 1 & 0\\ -s_{13} e^{i\delta} & 0 & c_{13} \end{pmatrix}
% \begin{pmatrix} c_{12} & s_{12} & 0\\ -s_{12} & c_{12} & 0\\ 0 & 0 & 1\end{pmatrix} ,
%\end{eqnarray}%%<<<<< eq: MNS -----
%where $c_{ij} \equiv \cos\theta_{ij}$and $s_{ij} \equiv \sin\theta_{ij}$.
%6
In our analysis, the following values~\cite{Abe:2014ugx,An:2013zwz,Aharmim:2011vm} obtained by neutrino oscillation measurements are used in order to search for a benchmark point of model parameters:
%%----- eq: nu params >>>>>
\begin{eqnarray}
%------------
%
&& \sin^2{2\theta_{23}}
=
 1 ,\quad
%
%------------
%
\sin^2{2\theta_{13}}
=
 0.09 ,\quad
%
%------------
%
\tan^2\theta_{12}
=
 0.427 ,\quad
%
%------------
%
\delta
=%&=&
 0 ,\quad
%
%------------
%
\bigl\{
 \alpha_1 ,\, \alpha_2 ,\, \alpha_3
\bigr\}
=
 \bigl\{
  0 ,\, 0 ,\, 0
 \bigr\} , \\
&& m_1
=
 10^{-4}\,\eV ,
\quad%\\
%
%------------
%
\Delta m^2_{21}
=%&=&
 7.46\times 10^{-5}\,\eV^2 ,\quad
%
%------------
%
\Delta m^2_{32}
=
 +2.51\times 10^{-3}\,\eV^2, \quad
 \text{where $\Delta m^2_{ij} \equiv m_i^2 - m_j^2$. } \ \ \
\end{eqnarray}
%%<<<<< eq: nu params -----

%%2
%1
 By using an ansatz~\cite{Kanemura:2014rpa} for the structure of Yukawa matrix $f_{\ell i}$, we found a benchmark point as
%%----- eq: benchmark >>>>>
\begin{eqnarray}
\!\!\!&&
f
=
 \begin{pmatrix}
  1.79
   & -2.49
   & -1.97
   & 2.56\\
  -1.82
   & 1.10
   & 1.30
   & -0.818\\
  1.40
   & -0.598
   & -0.905
   & 0.222
 \end{pmatrix} \times 10^{-2},
%f
%=
% \begin{pmatrix}
%  1.78686
%   & -2.48746
%   & -1.9737
%   & 2.55808\\
%%
%  -1.82223
%   & 1.10461
%   & 1.29624
%   & -0.818099\\
%%
%  1.40402
%   & -0.598335
%   & -0.904845
%   & 0.222417
% \end{pmatrix} \times 10^{-2},
\label{eq:f-benchmark}
%\\
%
%------------
%
\,\,%&&
h
=
 \begin{pmatrix}
  0.7
   & 0.8
   & 0.9
   & 1
 \end{pmatrix}^T,
%\\
%
 %------------
%
\,\,%&&
\bigl\{
 g_\BL^{} ,\, m_{Z^\prime}^{}
\bigr\}
=
 \bigl\{ 0.1 ,\, 4\,\TeV \bigr\}, 
\label{eq:BL-benchmark} 
\\
%
%------------
%
\!\!\!&&
\bigl\{
 m_{h^0}^{} ,\, m_{H^0}^{} ,\, \cos\theta_0
\bigr\}
=
 \bigl\{ 125\,\GeV ,\, 1\,\TeV ,\, 1 \bigr\} ,
\label{eq:neutral-benchmark}
%\\
%
%------------
%
\,\,%&&
\bigl\{
 m_{{\mathcal H}_1^0}^{} ,\, m_{{\mathcal H}_2^0}^{} ,\, \cos\theta_0^\prime
\bigr\}
=
 \bigl\{ 60\,\GeV ,\, 450\,\GeV ,\, 0.05 \bigr\} ,
\\
%
%------------
%
\!\!\!&&
m_{\eta^\pm}^{}
=
 420\,\GeV, 
%
%------------
%
\,\,%&&
(m_R)_1
=
 250\,\GeV,
%\\
%
%------------
%
\,\,%&&
\bigl\{
 m_{\psi_1}^{} ,\, m_{\psi_2}^{} ,\,
 m_{\psi_3}^{} ,\, m_{\psi_4}^{}
\bigr\}
=
 \bigl\{
  650\,\GeV ,\, 750\,\GeV ,\, 850\,\GeV ,\, 950\,\GeV
 \bigr\}. \qquad
\end{eqnarray}
%%<<<<< tab: benchmark -----
%The values of $\{ g_\BL^{} ,\, m_{Z^\prime}^{}\}$ mean $v_\sigma^{} = 60\,\TeV$. 
 %2
 The values of $\{ m_{h^0}^{} ,\, m_{H^0}^{} ,\, \cos\theta_0 \}$ correspond to $\lambda_\phi \simeq 0.13$, $\lambda_\sigma \simeq 2.8\times 10^{-4}$ and $\lambda_{\sigma\phi} = 0$. 
 %3
 The values of $\{ m_{{\mathcal H}_1^0}^{} ,\, m_{{\mathcal H}_2^0}^{} ,\, \cos\theta_0^\prime \}$ and $m_{\eta^\pm}^{}$ can be produced by $m_s^{} \simeq 60\,\GeV$, $m_\eta^{} \simeq 450\,\GeV$, $\mu_3^{} \simeq 57\,\GeV$ and $\lambda_{\eta\phi} \simeq 0.86$. 
%
%-----------------------------
%-----------------------------

%%%%%%%%%%%%%%%%%%%%%%%%%%%%%%%%%%
\subsection{Lepton flavor violation}

%The charged scalar $\eta^\pm$ contributes to the LFV decays of charged leptons. The formula for the branching ratio~(BR) of $\mu \to e\gamma$ can be calculated~\cite{Hisano:1995cp} as
We consider the condition of the LFV decays of charged leptons. 
The charged scalar $\eta^\pm$ contributes to the branching ratio~(BR) of $\mu \to e\gamma$ whose formula have been calculated~\cite{Hisano:1995cp}. 
% as
%%<<<<< eq: BR(meg) -----
%\begin{eqnarray} \BR(\mu\to e \gamma) = \frac{3\alpha_\EM^{}}{64\pi G_F^2}
% \left| \frac{1}{m_{\eta^{\pm}}^2} f_{\mu i}\, 
% F\!\left( \frac{ m_{\psi_i}^2 }{ m_{\eta^{\pm}}^2 } \right) (f^\dagger)_{ie} \right|^2 , \quad 
% \text{where \ \ } F\!\left(x\right) \equiv \frac{ 1 - 6x + 3x^2 + 2x^3 - 6x^2 \ln(x) } { 6\left( 1-x \right)^4 }. \label{eq:meg}\end{eqnarray}
%%<<<<< eq: BR(meg) -----
%
At the benchmark point, we have $\BR(\mu\to e\gamma) = 6.1\times 10^{-14}$
%$\BR(\mu\to e\gamma) = 6.06545\times 10^{-14}$
which satisfies the current constraint $\BR(\mu\to e\gamma) < 5.7\times 10^{-13}$~(90\%~C.L.)~\cite{Ref:MEG}.

%%%%%%%%%%%%%%%%%%%%%%%%%%%%%%%%%%
\subsection{Dark matter}

%1暗黒物質候補
In our model, the scalar ${\mathcal H}^0_1$ turns out to be the DM candidate due to the following reason. 
%%暗黒物質残存量(フェルミオン)
%2
If the DM is the fermion $\psi_1$, it annihilates into a pair of SM particles via the $s$-channel process mediated by $h^0$ and $H^0$.
%The cross section of the process is proportional to $\sin^2{2\theta_0}$.
%
%In order to obtain a sufficient annihilation cross section of $\psi_1$, a large mixing $\cos\theta_0 \simeq 1/\sqrt{2}$ is preferred~\cite{Ref:BL-2}.
%3
Even for a maximal mixing $\cos\theta_0 = 1/\sqrt{2}$~\cite{Ref:BL-2}, the observed abundance of the DM~\cite{Ade:2013zuv} requires $v_\sigma^{} \lesssim 10\,\TeV$.
%4 暗黒物質直接探索(フェルミオン)
The current constraint from direct searches of the DM~\cite{Ref:LUX} requires larger $v_\sigma^{}$ in order to suppress the $Z^\prime$ contribution. 
%\footnote{This is because $m_{Z^\prime}^{}/g_\BL$ is not $2 v_\sigma$ as usual but $2 v_\sigma/3$ in this model.}

%%暗黒物質残存量(スカラー)
%1
The scalar DM ${\mathcal H}^0_1$ at the benchmark point is dominantly made from $s^0$ which is a gauge-singlet field under the SM gauge group, because of the tiny mixing $\cos\theta_0^\prime = 0.05$.
%2
The annihilation of ${\mathcal H}^0_1$ into a pair of the SM particles is dominantly caused by the $s$-channel scalar mediation via $h^0$~\cite{Kanemura:2010sh} because $H^0$ is assumed to be heavy.
%3
The coupling constant $\lambda_{{\mathcal H}^0_1 {\mathcal H}^0_1 h^0}^{}$ for the $\lambda_{{\mathcal H}^0_1 {\mathcal H}^0_1 h^0}^{}\, v_\phi^{} {\mathcal H}^0_1 {\mathcal H}^{0\ast}_1 h^0$ interaction controls the annihilation cross section, the invisible decay $h^0 \to {\mathcal H}^0_1 {\mathcal H}^{0\ast}_1$ in the case of kinematically accessible, and the $h^0$ contribution to the spin-independent scattering cross section $\sigma_\SI$ on a nucleon.
%4
In Ref.~\cite{Cline:2013gha}, for example, we see that ${\mathcal H}^0_1$
with $m_{{\mathcal H}^0_1}^{} = 60\,\GeV$ and $\lambda_{{\mathcal H}^0_1 {\mathcal H}^0_1 h^0}^{} \sim 10^{-3}$ can satisfy constraints from the relic abundance of the DM and the invisible decay of $h^0$. 
%%暗黒物質直接探索(スカラー)
%1
We see also that the $h^0$ contribution to $\sigma_\SI$ is small enough to satisfy the current constraint $\sigma_\SI < 9.2\times 10^{-46}\,\text{cm}^2$ for $m_\DM^{} = 60\,\GeV$~\cite{Ref:LUX}.
%2
Although the scattering of ${\mathcal H}^0_1$ on a nucleon is mediated also by the $Z^\prime$ boson in this model, the contribution can be suppressed by taking a large $v_\sigma^{}$.
%3
The benchmark point corresponds to $v_\sigma^{} = 60\,\TeV$ and gives about $6.6\times 10^{-47}\,\text{cm}^2$ for the scattering cross section via $Z^\prime$, which is smaller than the current constraint~\cite{Ref:LUX} by an order of magnitude.
%4
Thus, the constraint from the direct search of the DM is also satisfied at the benchmark point.

%%%%%%%%%%%%%%%%%%%%%%%%%%%%%%%%%%
\subsection{Z' and $\nu_R$ search}

%%Z^\primeの生成
%1
The LEP-II bound $m_{Z^\prime}^{}/g_\BL^{} \gtrsim 7\,\TeV$~\cite{LEPZp} is satisfied at the benchmark point because of $m_{Z^\prime}^{}/g_\BL^{} = 40\,\TeV$ which we take for a sufficient suppression of $\sigma_\SI$ for the direct search of the DM.
%2
The production cross section of $Z^\prime$ with $g_\BL^{} = 0.1$ and $m_{Z^\prime}^{} = 4\,\TeV$ is about $0.3\,\text{fb}$ at the LHC with $\sqrt{s}=14\,\TeV$~\cite{Ref:BL-Pheno}. 
%\footnote{The production cross section becomes about $6\,\text{fb}$ if we take $g_\BL^{} = 0.05$ and $m_{Z^\prime}^{} = 2000\,\GeV$. 
%3
Notice that the current bound $m_{Z^\prime}^{} \gtrsim 3\,\TeV$ at the LHC~\cite{LHCZp} is for the case where the gauge coupling for $Z^\prime$ is the same as the one for $Z$, namely $g_\BL^{} \simeq 0.7$. 
%} 
%%Z^\prime崩壊
%1 Z^\primeの崩壊
Decay branching ratios of $Z^\prime$ are shown at the benchmark point in Table~\ref{tab:Zp-decay}. 
%2 (Z^\prime -> ) \psi_i の崩壊 ( -> \nu_R^{} {\mathcal H}^0_1)
Decays of $\psi_i$ are dominated by $\psi_i \to \nu_R^{} {\mathcal H}^0_1$
with the Yukawa coupling constants $h_{i1}$ because $y_{\ell i}^{}$ for $\psi_i \to \ell^\pm \eta^\mp$ are small in order to satisfy the $\mu \to e\gamma$ constraint.
%3 (Z^\prime -> ) {\mathcal H}^0_2の崩壊
The ${\mathcal H}^0_2$~($\simeq \eta^0$) decays into $h^0 {\mathcal H}^0_1$ via
the trilinear coupling constant $\mu_3^{}$. 
%4 (Z^\prime -> ) \eta^\pmの崩壊
The main decay mode of $\eta^\pm$ is $\eta^\pm \to W^\pm {\mathcal H}^0_1$
through the mixing $\theta_0^\prime$ between $\eta^0$ and $s^0$.

%%\nu_R^{}生成
%In this model, $\nu_R^{}$ is not the DM and can decay into the SM particles.
%Decay branching ratios for $\nu_R^{}$ are shown in Table~\ref{tab:nuR-decay}.
%1 \nu_R^{} -> H^0は禁止されている
The $\nu_R^{}$ decay into $H^0$ is forbidden because it is heavier than $\nu_R^{}$ at the benchmark point.
%2 \nu_R^{}の生成
Since the $\BL$ charge of $\nu_R^{}$ is rather small, $\nu_R^{}$ is not produced directly from $Z^\prime$.
%3 \psi_i -> \nu_R^{}
However, $\nu_R^{}$ can be produced through the decays of $\psi_i$.
%%\nu_R^{}崩壊
%1
As a result, about $18\,\%$ of $Z^\prime$ produces $\nu_R^{}$.
%2
For $\nu_R^{} \to W \ell$~($56\,\%$) followed by the hadronic decay of $W$~($68\,\%$), the $\nu_R^{}$ would be reconstructed.
%3
In this model, an invariant mass of a pair of the reconstructed $\nu_R^{}$ is not at $m_{Z^\prime}^{}$ in contrast with a naive model where only three $\nu_R^{}$ with $\BL = -1$
are introduced to the SM. 
%\footnote{In the naive model with $m_{Ra}^{} = 250\,\GeV$~(degenerate) and $m_{Z^\prime}^{} = 4\,\TeV$, the decay branching ratios of $Z^\prime$ into $\{ q\overline{q},\, \ell\overline{\ell},\, \nu_L^{}\overline{\nu_L^{}},\, \nu_R^{} \overline{\nu_R^{}} \}$ are $\{ 0.25,\, 0.38,\, 0.19,\, 0.19 \}$. }
%4
This feature of $\nu_R$ also enables us to distinguish this model from the previous model in Ref.~\cite{Ref:KNS} where $\nu_R$ with $\BL = 1$ can be directly produced by the $Z^\prime$ decay.

%--------------------------------
\begin{table}[t]
\begin{center}
%%---- tab: Z' decay >>>>>
\caption{
 Branching ratios of $Z^\prime$ decays.
}
\begin{tabular}{|c|c|c|c|c|c|c|c|c|c|c|}
   \hline
 \ \ \ $q\, \overline{q}$ \ \ \
 & \ \ \ $\ell\, \overline{\ell}$ \ \ \
 & \ $\nu_L^{} \overline{\nu_L^{}}$ \
 & \ $\nu_R^{} \overline{\nu_R^{}}$ \
 & \ $\psi_1 \overline{\psi_1}$ \
 & \ $\psi_2 \overline{\psi_2}$ \
 & \ $\psi_3 \overline{\psi_3}$ \
 & \ $\psi_4 \overline{\psi_4}$ \
 & \ ${\mathcal H}^0_1 {\mathcal H}^{0\ast}_1$ \
 & \ ${\mathcal H}^0_2 {\mathcal H}^{0\ast}_2$ \
 & \ $\eta^+ \eta^-$ \
\\ \noalign{\hrule height 1.8pt}
 $0.21$
 & $0.32$
 & $0.16$
 & $0.0059$
 & $0.046$
 & $0.045$
 & $0.044$
 & $0.043$
 & $0.041$
 & $0.038$
 & $0.039$
   \\ \hline
\end{tabular}
\label{tab:Zp-decay}
%%<<<<< tab: Z' decay -----
%
%%---- tab: nuR decay >>>>>
%\caption{Branching ratios of $\nu_R^{}$ decays.}
%\begin{tabular}{|c|c|c|c|} \hline
% \ \ \ $W^+ \ell^- + W^- \ell^+$ \ \ \
% & \ \ \ $Z \nu_L^{} + Z \overline{\nu_L^{}}$ \ \ \
% & \ $h^0 \nu_L^{} + h^0 \overline{\nu_L^{}}$ \
% & \ $H^0 \nu_L^{} + H^0 \overline{\nu_L^{}}$ \ \\ \noalign{\hrule height 1.8pt}
% $0.56$ & $0.28$ & $0.16$ & $0$ \\    \hline \end{tabular}
%\label{tab:nuR-decay}
%%<<<<< tab: nuR decay -----
\end{center}
\end{table}
%--------------------------------

%%%%%%%%%%%%%%%%%%%%%%%%%%%%%%%%%%
%%%%%%%%%%%%%%%%%%%%%%%%%%%%%%%%%%
%%%%%%%%%%%%%%%%%%%%%%%%%%%%%%%%%%
%%%%%%%%%%%%%%%%%%%%%%%%%%%%%%%%%%
\section{Conclusions}

%%
%1 先行研究の改善
We have improved the model in Ref.~\cite{Ref:KNS} by considering anomaly cancellation of the $\U(1)_\BL$ gauge symmetry.
%2 4つのアノマリーの場合分け
We have shown that there are four anomaly-free cases of $\BL$ charge assignment, and three of them have an unbroken global $\U(1)_\DM$ symmetry. %(one of the three is not acceptable because two neutrinos become massless). 
%%
%1 暗黒物質の安定性
The $\U(1)_\DM$ guarantees that the lightest $\U(1)_\DM$-charged particle is stable such that it can be regarded as a DM candidate. 
%2 B-Lの破れから新粒子質量
The spontaneous breaking of the $\U(1)_\BL$ symmetry generates 
%new fermion mass terms which do not exist in the SM; namely, the Dirac mass term of neutrinos, 
the Majorana mass term of $\nu_R^{}$ %, 
and masses of new fermions $\psi$. 
%3 1ループDirac質量 Especially, 
In addition, the Dirac mass term of neutrinos is generated at the one-loop level where the DM candidate involved in the loop. 
%4 2ループニュートリノ質量
Tiny neutrino masses are obtained at the two-loop level. 

%%
%1 フェルミオン暗黒物質は排除される
The case of the fermion DM is excluded, and the lightest $\U(1)_\DM$-charged scalar ${\mathcal H}_1^0$ should be the DM in this model.
%2 実験を満たすベンチマークを示した
We have found a benchmark point of model parameters which satisfies current constraints from neutrino oscillation data, lepton flavor violation searches, the relic abundance of the DM, direct searches for the DM, and the LHC experiments.
%
%By virtue of the radiative mechanism for the Dirac mass term of neutrinos, very heavy $\nu_R^{}$ are not required for tiny neutrino masses. Therefore, 
%3
In such radiative seesaw models, $\nu_R^{}$ would be produced at the LHC\@.
%4 普通のB-Lとは違う
%In contrast to a naive model where three $\nu_R^{}$ have $\BL = -1$ and the model in Ref.~\cite{Ref:KNS} where $\nu_R^{}$ have $\BL = 1$, the $\nu_R^{}$ with $\BL = -1/3$ in this model cannot be directly produced by the $Z^\prime$ decay, but can be produced by the cascade decay $Z^\prime \to \psi_i \overline{\psi}_i \to \nu_R^{} \overline{\nu_R^{}} {\mathcal H}_1^0 {\mathcal H}_1^{0\ast}$.
In our model, $\nu_R^{}$ cannot be directly produced by the $Z^\prime$ decay, but can be produced by the cascade decay $Z^\prime \to \psi_i \overline{\psi}_i \to \nu_R^{} \overline{\nu_R^{}} {\mathcal H}_1^0 {\mathcal H}_1^{0\ast}$.
%5 右巻きニュートリンの検証
%The invariant mass distribution of $\nu_R^{} \overline{\nu_R^{}}$ does not take a peak at $m_{Z^\prime}^{}$, which could be a characteristic signal of this kind of models with the unusual $\BL$ charge of $\nu_R^{}$.
By the unusual $\BL$ charge of $\nu_R^{}$, the invariant mass distribution of $\nu_R^{} \overline{\nu_R^{}}$ does not take a peak at $m_{Z^\prime}^{}$, which could be a characteristic signal.

\begin{acknowledgments}
This work is based on the collaboration with Shinya Kanemura and Hiroaki Sugiyama. 
I would like to thank them for their support.
\end{acknowledgments}

\bigskip % extra skip inserted

\end{document}